\documentclass[aps,pra,twocolumn,showpacs,floatfix]{revtex4}
\usepackage{epsfig}
\usepackage{graphicx}
\usepackage{dcolumn}
\usepackage{amsthm,amsmath}

\begin{document}

\title{Emending thermal dispersion interactions of Li, Na, K and Rb alkali metal-atoms with graphene in the Dirac model}

\author{Kiranpreet Kaur, Jasmeet Kaur, and Bindiya Arora \footnote{Email: arorabindiya@gmail.com}}
\affiliation{Department of Physics, Guru Nanak Dev University, Amritsar, Punjab-143005, India}
\author{B. K. Sahoo \footnote{Email: bijaya@prl.res.in}}
\affiliation{Theoretical Physics Division, Physical Research Laboratory, Navrangpura, Ahmedabad-380009, India}
\date{Recieved Date; Accepted Date}

\begin{abstract}
Using accurate dynamic polarizabilities of Li, Na, K and, Rb atoms, we scrutinize the  
thermal Casimir-Polder interactions of these atoms with a single layered graphene. Considering
the modified Lifshitz theory for material interactions, we reanalyze the dispersion coefficients 
($C_3$s) of the above atoms with graphene as functions of separation distance, gap parameter and 
temperature among which some of them were earlier studied by estimating dynamic polarizabilities of the 
above atoms from the single oscillator model approximation. All these $C_3$ coefficients have been
evaluated in the framework of the Dirac model. The interactions are described for a wide range of 
distances and temperatures to demonstrate the changes in behavior with the varying conditions of the 
system and also sensitivities in the interactions are analyzed by calculating them for different values
of the gap parameter. From these analyses, we find a suitable value of the gap parameter for which
the true nature of the interactions in graphene can be surmised more accurately.
\end{abstract}
\pacs{73.22.Pr, 78.67.-n, 12.20.Ds}
\maketitle

\section{Introduction}\label{sec1}
Owing to unique electronic, optical, mechanical, thermal and magnetic properties of carbon nano
structures \cite{friedrich, intro2}, investigating interactions of one of its contenders, 
graphene having mono layer carbon atoms, with other materials such as atoms, have 
drawn much attentions in both scientific and industrial researches in recent times 
\cite{intro2,nature}. On the other hand, atoms belonging to the alkali group are the favorites 
among the experimentalists to carry out studies either on the scattering phenomena or to investigate 
fundamental principles of the governing interactions interplaying within these systems. Since the 
structures of these atoms are well understood to some extent by now, hence it is possible to 
manipulate their interactions and control many of their systematics in the course of the experiments 
involving these atoms. The dispersion forces acting between the materials are described by dividing 
the entire interaction regime into two parts: non-retarded and retarded distances. In the non-retarded 
regime (at small distances), role of the speed of light is neglected and the interactions are generally 
explained using the van der Waals forces ~\cite{vander,dirac-hydro}. On the other hand, the separation 
distances in the retarded regime are compared to the speed of light times a characteristic time and the 
interactions are usually due to the Casimir-Polder forces \cite{casimir1,blagov}. A better perception 
about the underlying physics involved in these interactions can lay out many applications in the upcoming 
nano-technologies; especially in the silicon integrated circuit technology based micro-electromechanical 
(MEM) and nano-electromechanical (NEM) systems. Particularly, the Casimir-Polder forces, due to their 
strong distance dependencies, can produce large pulls-in and cause stictions in the MEM and NEM devices 
~\cite{harjeet1}. Other such pronounced applications include fabricating hydrogen storage  devices 
~\cite{intro3,intro4,intro5,harjeet}, explaining certain physical, chemical and biological processes 
~\cite{intro6,intro7,intro8,intro9}, etc. For example, a better understanding of the interactions of the 
Li atoms with graphene is helpful for designing a better storage mechanism for the hydrogen gas 
~\cite{ataca,du,harjeet}, to construct high-quality superconductors \cite{castro,intro16}, in the development
of the sophisticated up-gradation technologies for the Li-ion batteries ~\cite{konga}, etc. Experimental 
investigations of these interactions are extremely complicated. Even though many theoretical approaches, 
such as density functional theory ~\cite{dino,dft1,dft4,dft2,dft3}, lower-order many-body 
perturbation theory ~\cite{mbpt1}, Lifshitz theory \cite{caride,som1} etc., have been employed to describe 
these interactions, but they are not so facile for studying these interactions \cite{som1,lifshitzbook,
lifshitz1,dazy}. Within the Lifshitz theory, two models are generally acceptable to explain these interactions 
which are known as the hydrodynamic and Dirac models \cite{dirac-hydro2}. Among these two, the Dirac model is 
more famous on the physical ground in which the quasi-particle fermion excitations in the graphene are treated 
as massless Dirac fermions moving with the Fermi velocities. 

Another important attribute to study the interactions of graphene with atoms lies in the rigorous treatment of 
the electron correlation effects to calculate the properties of the involved atoms accurately.  In a recent work, 
we had investigated the role of using accurate values of the dynamic polarizabilities of the alkali atoms to 
describe these interactions both by the hydrodynamic and Dirac models at zero temperature \cite{bindiya3}. We had
observed in that study that the $C_3$ coefficients change significantly in the heavier systems, like in the K and 
Rb atoms, when accurate polarizability values of the atoms are used. Since zero temperature condition is not a
realistic situation for the practical applications, in this work, we intend to find out the role of the accurate 
values of the dynamic polarizabilities of the atoms in the behavioral investigations of the graphene-atom 
interactions at finite temperatures, including the room temperature, and compare them with the previously obtained 
results considering the dynamic polarizabilities from the single oscillator model (SOM) \cite{caride, dirac-hydro2}. In addition, 
we also make an attempt to identify a regime in which it would be possible to make a better comparison between 
the theoretical and experimental potentials and a rational value of the mass gap parameter for graphene can be 
extracted to describe the graphene-atom interaction potentials shrewdly. Unless stated otherwise, we use 
atomic unit (au) through out the paper.

\section{Theory}\label{sec2}
The general expression of van der Waals and Casimir Polder energy for an atom with graphene, separated by distance $a$, 
is expressed in terms of dispersion coefficients as ~\cite{dirac-hydro}
\begin{eqnarray}
E(a)= - \frac{C_3}{a^3},
\end{eqnarray}
where the dispersion coefficient $C_3$  at zero temperature is defined as
\begin{eqnarray}
C_3(a)&=&-\frac{1}{16\pi}\int_0^{\infty}d\xi\alpha(\iota\xi)\int_{2a\xi \alpha_{fs} }^{\infty}dye^{-y}y^2 \nonumber\\
 & &\left(2r_{{TM}}-\frac{4a^2 \alpha_{fs}^2\xi^2}{y^2}(r_{{TM}}+r_{{TE}})\right),
\end{eqnarray}
with ${r_{TM}}$ and ${r_{TE}}$ as the Fresnel reflection coefficients of the electromagnetic oscillations on 
graphene for the transverse components of the electromagnetic field, respectively, which are given by
\begin{eqnarray}
r_{TM}(\iota \xi, k_{\perp})= \frac{\epsilon(\iota \xi) q(\iota \xi, k_{\perp})- k(\iota \xi, k_{\perp})}{ \epsilon(\iota \xi) q(\iota \xi, k_{\perp})+ k(\iota \xi, k_{\perp})}
\end{eqnarray}
and
\begin{eqnarray}
r_{TE}(\iota \xi, k_{\perp})= \frac{ q(\iota \xi, k_{\perp})- k(\iota \xi, k_{\perp})}{ q(\iota \xi, k_{\perp})+ k(\iota \xi, k_{\perp})} .
\end{eqnarray}
In these expressions, $k_{\perp} \equiv (k_x,k_y)$ are the components of wave number $k$ of the electromagnetic field,
parameter $q \equiv q(\iota \xi)= \sqrt{ k_{\perp}^2+ \alpha_{fs}^2 \xi^2}$ with the fine structure constant 
$\alpha_{fs}$ and $\epsilon(\iota, \xi)$ is the dynamic dielectric permittivity of graphene with the imaginary frequency and is 
related to $k$ as $k(\iota \xi)= \sqrt{ k_{\perp}^2+ \epsilon(\iota \xi) \alpha_{fs}^2 \xi^2}$. Appearance of the imaginary 
frequencies in the above expressions reveal that only virtual electronic excitations are associated with the 
polarization during the interactions and none of the energies get transferred between the objects.

In the practical applications, these interactions are carried out at finite temperature, mostly at the room temperature. 
For this purpose, the generalized expression for the Casimir-Polder energy at a finite temperature $T$ is obtained by 
replacing the integral over frequency to sum over the Matsubara frequencies as~\cite{harjeet1}
\begin{equation}
\int_0^\infty \frac{d\xi}{2\pi}\rightarrow\frac{1}{\beta}{\sum_{n=0}^\infty},
\end{equation}
where $\beta=\frac{1}{k_BT}$ with the Boltzmann constant $k_B$. Therefore, the general expression for the $C_3$ 
coefficient in terms of the reflection coefficients $r_{TM}$ and $r_{TE}$ is given by~\cite{dirac-hydro2} 
\begin{eqnarray}
C_3(a,T)&=&- {\frac{k_B T}{8}} {\sum'_l} \alpha(\iota \zeta_l \omega_c) \int_{\zeta_l}^{\infty}dy \{ e^{-y} 2y^2 {\zeta_l}^2 \nonumber \\
&& r_{\rm{TM}}(\iota\zeta_l,y) \left[r_{\rm{TM}}(\iota\zeta_l,y)+r_{\rm{TE}}(\iota\zeta_l,y)\right]\}.\label{eq-u}
\end{eqnarray} 
Here, it has been pretended that graphene is in thermal equilibrium at temperature  $T$, the dynamic 
polarizability $\alpha(\iota{\xi_l})$ of the atom can be calculated along the imaginary Matsubara frequencies 
${\xi_l}=2 \pi k_B T l/{\hbar}$ with $l=0, 1, 2,..,$ and $\zeta_l=(\xi_l /{\omega_c})$ for the dimensionless 
Matsubara frequencies with the character frequency $\omega_c=1/(2a \alpha_{fs})$. The prime over the summation sign 
indicates multiplication by a factor $1/2$ in the $l=0$ term. 
 
 The reflection coefficients of the electromagnetic oscillations on graphene can be determined using either the 
hydrodynamic model ~\cite{barton,barton1,bordag1} or Dirac model ~\cite{semen,dvince,dft3,drosdoff,dr1,dr2,dr3}. 
In the hydrodynamic model, graphene is considered as an infinitesimally thin positively charged flat sheet 
carrying a homogeneous fluid with some mass and negative charge densities. This model, however, does not take into
account some of the important properties of the graphene which are important at the low energies; specifically that 
the energies of the quasi-particles of mass $m$, introduced within this model, are linear functions of the momentum. 
On the other hand, in case of the Dirac model, the dispersion relations are linear at any energy value. Hence, on 
the physical ground the Dirac model is more acceptable and has been considered in the present work. In this 
model, the reflection coefficients are given in terms of the components of dimensionless polarization tensors
$\tilde{\Pi}_{00}$ and $\tilde{\Pi}_{\rm tr}$ as \cite{dirac-hydro2,ivf}
\begin{eqnarray}
r_{\rm{TM}}&=&\frac{y{\tilde{\Pi}}_{00}}{y{\tilde{\Pi}}_{00}+2(y^2-\zeta_l^2)}
\end{eqnarray}
and
\begin{eqnarray}
r_{\rm{TE}}&=&-\frac{(y^2-\zeta_l^2)\tilde{\Pi}_{tr}-y^2\tilde{\Pi}_{00}}{(y^2-\zeta_l^2)(\tilde{\Pi}_{tr}+2y)-y^2\tilde{\Pi}_{00}},~\label{eq-dirac}
\end{eqnarray} 
where $\tilde{\Pi}_{00,\rm{tr}}$ is related with $\Pi_{00,\rm{tr}}$ as ${\tilde{\Pi}}_{00,\rm{tr}}(\iota\zeta_l,y)=(2a/\hbar)\Pi_{00,\rm{tr}}(\iota\zeta_l,y)$.

The expressions for the components of the polarization operators at the non-zero temperatures are explicitly given
by \cite{dirac-hydro, dirac-hydro2}  
\begin{eqnarray}
\tilde{\Pi}_{00}(\iota \zeta_l,y)&=& 8 \alpha(y^2-\zeta_l^2)\int_0^1 dx\frac{x(1-x)}{\left[\Delta^2+x(1-x)f(\zeta_l,y)\right]^{1/2}}\nonumber\\
&+&\frac{8\alpha}{{\tilde{v}_F}^2}\int_0^1 dx\left\lbrace \frac{\tau}{2\pi}{\rm{ln}}(1+2\cos (2\pi lx)e^{-g(\tau,\zeta_l,y)}\right.\nonumber\\
&+& e^{-2g(\tau,\zeta_l,y)})-\frac{\zeta_l}{2}(1-2x)\nonumber\\
& &\frac{\sin (2\pi lx)}{\cosh g(\tau,\zeta_l,y)+\cos (2\pi lx)}\nonumber\\
&+&\frac{{\tilde{\Delta}}^2+\zeta_l^2x(1-x)}{{\left[{\tilde{\Delta}}^2+x(1-x)f(\zeta_l,y)\right]}^{1/2}}\nonumber\\
& &\left.\frac{\cos (2\pi lx)+e^{-g(\tau,\zeta_l,y)}}{\cosh g(\tau,\zeta_l,y)+\cos(2\pi lx)}\right\rbrace, \label{eq-pi00}
\end{eqnarray}
where $\Delta$ is known as the gap parameter which is introduced to regularize the theory and $\tau=4\pi \alpha_{fc} k_BT/\hbar$. 
In the above expression, few parameters are introduced such as
$\tilde{v}_F= \alpha_{fs} v_F$ with the Fermi velocity $v_F$ and the dimensionless constants as $\tilde{\Delta}=\Delta/(\hbar \omega_c)$,
$f(\zeta_l,y)$ and $g(\tau, \zeta_l,y)$. Although the exact value of the $\Delta$ parameter depends on the interaction
strength and range, its maximum value is often assumed to be 0.1 eV \cite{dirac-hydro,dirac-hydro2}. However, for a 
pristine (gapless) graphene, $\Delta=0$ is meaningful as in this case the mass of the quasi-particle $m=0$.
For the chemical potential $\mu$ to be zero, $f$ and $g$ are given by
\begin{eqnarray}
f(\zeta_l,y)&=&{\tilde{v}_F}^2 y^2+(1-{\tilde{v}_F}^2)\zeta_l^2
\end{eqnarray}
and
\begin{eqnarray}
g(\tau,\zeta_l,y)&=&\frac{2\pi}{\tau}{\left[{\tilde{\Delta}}^2+x(1-x)f(\zeta_l,y)\right]}^{1/2}. 
\label{eq-fg}
\end{eqnarray}

The polarization tensor ${\tilde{\Pi}_{tr}}$ is defined in terms of the above dimensionless variables as
{\small
\begin{eqnarray}
{\tilde{\Pi}}_{tr}(\iota \zeta_l,y)&=& 8 \alpha\left[y^2+f(\zeta_l,y)\right]\int_0^1 dx\frac{x(1-x)}{\left[\Delta^2+x(1-x)f(\zeta_l,y)\right]^{1/2}}\nonumber\\
&+&\frac{8\alpha}{{\tilde{v}_F}^2}\int_0^1 dx\left\lbrace \frac{\tau}{2\pi}{\rm{ln}}(1+2 \cos (2\pi lx)e^{-g(\tau,\zeta_l,y)}.\right.\nonumber\\
&+&e^{-2g(\tau,\zeta_l,y)})-\frac{\zeta_l\left(1-2{\tilde{v}_F}^2\right)}{2}(1-2x)\nonumber\\
& &\frac{\sin (2\pi lx)}{\cosh g(\tau,\zeta_l,y)+\cos (2\pi lx)}\nonumber\\
&+&\frac{{\tilde{\Delta}}^2+x(1-x)\left[(1-\tilde{v}_F^2)^2\zeta_l^2-\tilde{v}_F^4 y^2\right]}{{\left[{\tilde{\Delta}}^2+x(1-x)f(\zeta_l,y)\right]}^{1/2}}\nonumber\\
& &\left.\frac{\cos (2\pi lx)+e^{-g(\tau,\zeta_l,y)}}{\cosh g(\tau,\zeta_l,y)+\cos (2\pi lx)}\right\rbrace.
~\label{eq-pitr}
\end{eqnarray}}

By setting $T=0$ in the above formulas, the polarization operators reduce to the following forms \cite{dirac-hydro,dirac-hydro2}
\begin{eqnarray}
{\tilde{\Pi}}_{00}(\iota\zeta,y)&=&\alpha\frac{y^2-\zeta^2}{f(\zeta,y)}{\tilde{\Phi}}_{00}(\iota\zeta,y)
\end{eqnarray}
and
\begin{eqnarray}
{\tilde{\Pi}}_{tr}(\iota\zeta,y)&=&\alpha\frac{y^2+f(\zeta,y)}{f(\zeta,y)}{\tilde{\Phi}}_{00}(\iota\zeta,y),\label{eq-t01}
\end{eqnarray}
where $\zeta$ is the continuous dimensionless frequency and
\begin{eqnarray}
{\tilde{\Phi}}_{00}(\iota\zeta,y)&=&4{\tilde{\Delta}}+2\sqrt{f(\zeta,y)}\left[1-4\frac{{\tilde{\Delta}}^2}{f(\zeta,y)}\right] \nonumber \\ && {\rm{\arctan}}\frac{\sqrt{f(\zeta,y)}}{2{\tilde{\Delta}}}, \label{eq-t02}
\end{eqnarray}
which leads to the following expressions for the reflection coefficients at zero temperature
\begin{eqnarray}
r_{\rm{TM}}(\iota\zeta,y)&=&\frac{\alpha y{\tilde{\Phi}_{00}}(\zeta,y)}{\alpha y{\tilde{\Phi}_{00}}(\zeta,y)+2f(\zeta,y)}
\end{eqnarray}
and
\begin{eqnarray}
r_{\rm{TE}}(\iota\zeta,y)&=&-\frac{\alpha{\tilde{\Phi}_{00}}(\zeta,y)}{\alpha{\tilde{\Phi}_{00}}(\zeta,y)+2y}. 
\label{eq-t03}
\end{eqnarray}

\section{Dynamic Polarizability}\label{sec3}

The dynamic dipole polarizability of an alkali metal atom in its ground state $|\Psi_n \rangle$ at the
imaginary frequency ($\iota \omega$) is given by
\begin{eqnarray}
\alpha(\iota \omega)&=& \sum_{I \ne n}  \frac{ (E_I -E_n) |\langle \Psi_n | D | \Psi_I\rangle |^2}{(E_I - E_n)^2 + \omega^2} \nonumber \\ 
              &=& \frac{2}{3(2J_n+1)} \sum_I \frac{ (E_I-E_n) |\langle \Psi_n || D || \Psi_I\rangle |^2}{(E_I - E_n)^2 + \omega^2}
\end{eqnarray}
where the subscripts $n$ and $I$ are for the ground and intermediate states  and $J_n$  is the total angular momentum 
of the ground state, $E$s are the energies of the states  and 
$\langle \Psi_n||D||\Psi_I \rangle$ is the reduced matrix element of the electric dipole (E1) operator $D$ between 
the ground state and the intermediate state.

In case, a sufficiently large number of intermediate states $|\Psi_I \rangle$ are known which can predominantly 
contribute in the determination of $\alpha$, then the above expression is very convenient to calculate the dynamic 
polarizabilities for any value of $\omega$ by just calculating the reduced E1 matrix elements of those known states 
and their corresponding excitation energies. In fact, it can also leverage the accuracies of the results by replacing 
the best known E1 matrix elements and energies either from the precise measurements or precise calculations from 
the potential many-body methods. We take liberty to adopt this approach for the accurate determination of the dynamic polarizabilities of the 
alkali atoms. Owing to the fact that many of the low-lying states of the alkali atoms can be expressed by a valence 
orbital attached to a common core, all these states have been well studied using a variety of many-body methods 
including the all order relativistic coupled-cluster (RCC) methods \cite{bindiya1,bindiya2,sahoo1,clark,harjeet}. Also, a sufficient number of 
transition properties of these systems are experimentally observed \cite{bindiya1,sahoo1,volz}. As a result, it is commended to 
make use of these quantities for precise estimations of the polarizabilities in these atoms. On the entrust of 
obtaining high precision dipole polarizabilities with the inferences of these known quantities, we have tabulated 
the most precise E1 matrix elements for a large number of transitions in our earlier work \cite{harjeet}. Along with 
the contributions from the above matrix elements, the other contributions from the continuum and corrections from 
the core and core-valence correlations are required to accomplish the final results for the polarizabilities. Since 
these contributions are relatively small, they are estimated using lower order methods as have been described in 
detail in \cite{harjeet}.

\begin{figure}[t]
\includegraphics[scale=0.68]{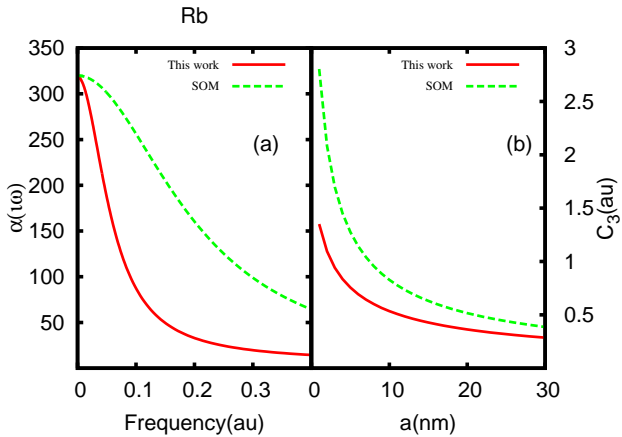}
\caption{(Color online) (a) Dynamic polarizabilities and (b) $C_3$ coefficients of the Rb atom interacting with 
the graphene layer at $T=300^{\circ}$ K, as functions of frequency and separation distance respectively, are shown 
from the present calculations and other works obtained using SOM \cite{dirac-hydro2}. }~\label{som}
\end{figure}

\begin{table}[t]
\caption{Comparison of our static polarizabilities (in au) of the ground states of the Li, Na, K and Rb alkali atoms
with the experimental results and values used in Ref. \cite{dirac-hydro2} for SOM.}
\begin{ruledtabular}
\begin{tabular}{l c c c c}
Atom			& Li & Na & K & Rb\\
\hline 
$\alpha(0)$ & & \\
Present	& 164.05 & 162.32 & 289.72 & 318.47\\
Ref.~\cite{dirac-hydro2} &  & 162.7(8) &  & 319.9(6.1) \\
Experiment  & 164.2(11)$^a$ & 162.7(8)$^b$  & 290.58(1.42)$^c$ & 318.79(1.42)$^c$
\end{tabular}
\end{ruledtabular}
$^a$Ref.~\cite{miffre}, $^b$Ref.~\cite{ekstrom}, $^c$Ref.~\cite{holmgren}.
\end{table}

As has been mentioned earlier, some of the previous works estimate the dynamic polarizabilities of the alkali atoms
for the required analysis using SOM  ~\cite{caride, som1, dirac-hydro2} in which the expression for the
dynamic polarizability is given by
\begin{equation}
 \alpha(\iota \omega_c \zeta_l)=\frac{\alpha(0)}{1+(\omega_c^{2}/\omega_0^{2})\zeta_l^{2}}, 
 ~\label{eq-pol}
\end{equation}
where $\alpha(0)$ is the static polarizability and $\omega_0$ is the characteristic absorption frequency of an 
alkali metal atom. Evidently, this is a bruteforce approach to acquire the dynamic polarizabilities at any frequency
when the $\alpha(0)$ and $\omega_0$ values of the atom are known.

\begin{figure}[t]
\includegraphics[scale=0.68]{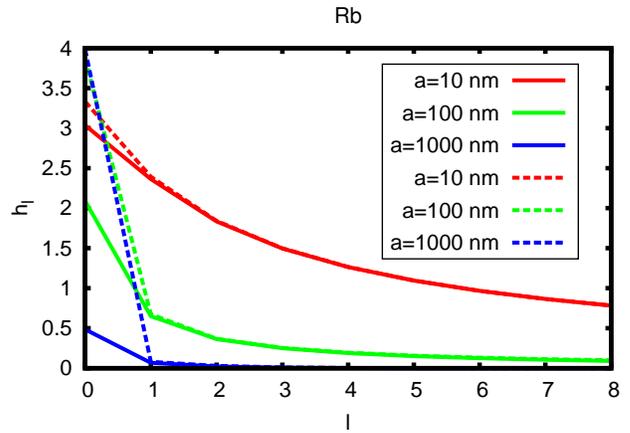}
\caption{(Color online) Variation of integral $h(l)$ as a function of $l$ using the reflection coefficients at $a=$ 10 
(red), 100 (green) and 1000 (blue) nm. The dashed and solid lines correspond to the values obtained using the 
reflections coefficients at the temperatures $T=0^{\circ}$ K and $300^{\circ}$ K, respectively.}~\label{rtme}
\end{figure}

\section{Results and Discussion}\label{sec7}

 In Table \ref{som}, we present the static polarizabilities that are reported by us \cite{harjeet} and compare them
with the results that are used in the earlier works from SOM \cite{dirac-hydro2} and the experimental results
\cite{ekstrom,miffre,holmgren}. The calculation details of our polarizability results are explained in \cite{harjeet}
and in the references therein. In contrast to the procedure for obtaining the dynamic polarizabilities of the atoms 
using SOM, our calculations can provide these results for both the static and dynamic polarizabilities at the same 
levels of accuracies. To outline the procedure followed in our calculations, the principal E1 matrix elements are 
obtained from the measurements of the lifetimes of the low-lying states of the considered atoms. Other important E1 
matrix elements are obtained using the RCC method, among which accuracies of some of the matrix elements obtained by
the RCC method are further ameliorated by trying to reproduce the experimental results of the scalar 
polarizabilities of the excited atomic states using these matrix elements. Excitation energies from the national 
institute for standards and technology (NIST) were used in order to avoid uncertainties arising from the theoretical
calculations. In the above mentioned SOM calculations, $\omega_0$ values for the Na and Rb atoms were taken as 
2.14 and 5.46 eV, respectively. To demonstrate the differences arising in the dynamic polarizability values from 
both the calculations, we consider the Rb atom as an example and plot these values from our calculations and those 
from SOM used in the above earlier works against frequencies (in au) in Fig.~\ref{som}. As seen in the figure, the 
single oscillator model values differ significantly from our results. From the comparisons between the measurements 
and the calculated results, as given in Table \ref{som}, 
it is obvious that our static polarizabilities agree well with the experimental values and are also more precise, and we expect the same precision in our dynamic polarizabilities over the previously used dynamic polarizabilities.
This suggests that the $C_3$ results
that are going to be evaluated in the present work are naturally going to be more reliable than the previously 
estimated results and the interaction potentials between the considered alkali atoms and the graphene can be 
apprehended better.

 During our computations, we noted that Eq.~(\ref{eq-dirac}) fails at distances greater than $30$ nm owing to the 
fact that for some particular combinations of `$a$' and `$l$', the expression for $r_{\rm{TE}}$ almost diverges 
leading to unphysical outcomes. For instance at $a=$ 36 nm, $l=$ 59 and $y \approx 3.51$, the denominator of 
$r_{\rm{TE}}$ is nearly equal to zero. Thus, it is concluded that for the large distances, especially when $l>30$ nm,
the Dirac model might not be giving appropriate expressions to describe the interactions. This steers to look into 
some alternative approach to deal with the above situation in which the reflection coefficients for the graphene 
under the thermal conditions can be admissible. The above problem to determine the $C_3$ coefficients in our 
calculations is vanquished in the following way. Instead of using the thermal reflection coefficients for all the $l$ 
components in Eq. (\ref{eq-u}), this is simplified by evaluating the thermal Eqs. (\ref{eq-dirac}), (\ref{eq-pi00}), 
(\ref{eq-fg}) and (\ref{eq-pitr}) only for the $l=0$ term and non-thermal
Eqs. (\ref{eq-t01}), (\ref{eq-t02}) and (\ref{eq-t03}) are evaluated for the $l>0$ terms while determining the 
reflection coefficients. This can be justified by plotting the integral, $h(l)= \int_{\zeta_l}^{\infty}dy e^{-y}
\{2y^2 r_{\rm{TM}}(\iota\zeta_l,y)-{\zeta_l}^2\left[r_{\rm{TM}}(\iota\zeta_l,y)+r_{\rm{TE}}(\iota\zeta_l,y)\right]\}$, 
inside the summation of Eq.(\ref{eq-u}) in Fig. \ref{rtme} by substituting the corresponding reflection coefficients
for the temperature at $T=0^{\circ}$ K and at $T=300^{\circ}$ K as a function of $l$, which are shown in the dashed 
and solid lines, respectively. As seen from the graph, the use of $l>0$ terms in the evaluation of the $h(l)$ 
function at $T=0^{\circ}$ K temperature leads to almost the same value of $h(l)$ as in the case of the temperature 
at $T=300^{\circ}$ K. In fact, this was extensively analyzed in Ref. \cite{dirac-hydro2}, which is further supported
by our findings and it justifies to consider the above mentioned assumptions in the determination of the reflection 
coefficients at the non-zero thermal conditions. Therefore, it has to be noted that the $C_3$ coefficients which are 
evaluated below are under these conjectures. 

 In the foregoing sub-sections, we discuss the interactions as the functions of the separation distance, gap 
 parameter and temperature of the system.

\subsection{$C_3$ as a function of separation distance}

\begin{figure}
\includegraphics[scale=0.68]{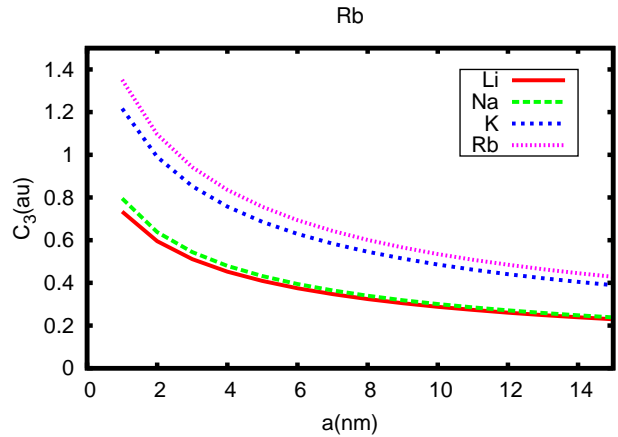}
\caption{(Color online) The $C_3$ coefficients (in au) as function of the atom-graphene separation distance for the 
alkali metal atoms Li, Na, K and, Rb, interacting at $T=300^{\circ}$ K.}~\label{c3a}
\end{figure}

In Fig. \ref{c3a}, we show the graph between the $C_3$ coefficients and the separation distance $a$ (in nm) for the 
Li (solid red curve), Na (long dashed green curve), K (short dashed blue curve) and Rb (dotted pink curve) atoms 
interacting with a graphene layer at the room temperature $T=300^{\circ}$ K and with the gap parameter 
$\Delta = 0.01$ eV. As was expected, the magnitudes of the interactions for the bigger atoms, say Rb, are 
found to be larger than the smaller atoms, say Li. It can be observed from the figure that the interactions 
between the atoms and the graphene layer are negligibly small at the large separation distances, whilst these are 
very effective at the smaller separation distances. These behaviors are in agreement with the findings of Ref.
\cite{dirac-hydro2} for the Na and Rb atoms interacting with graphene in the Dirac model, but our $C_3$ values 
are presumed to be more accurate than the given coefficients in \cite{dirac-hydro2} due to the use of the accurate 
dynamic polarizabilities of the considered atoms.

\subsection{$C_3$ as a function of gap parameter}

\begin{figure}[t]
\includegraphics[scale=0.67]{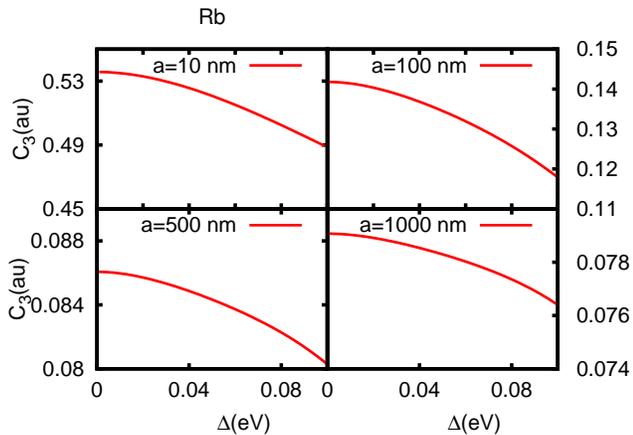}
\caption{(Color online) The $C_3$ coefficients (in au) of the alkali metal Rb atom as function of gap parameter
$\Delta$ (in eV) for four different values of the separation distance, $a=$10, 100, 500, and 1000 nm (clockwise), 
with $\Delta=0.01$ eV.}~\label{c3deltaa}
\end{figure}

\begin{figure}[t]
\includegraphics[scale=0.67]{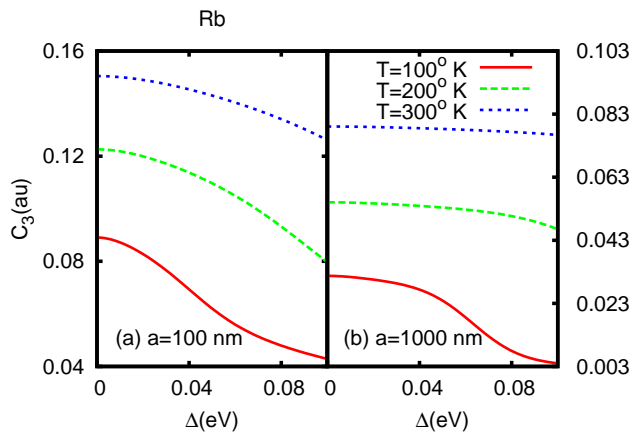}
\caption{(Color online) The $C_3$ coefficients of alkali metal Rb atom as function of gap parameter $\Delta$ 
(in eV) for three different values of the temperature, $T=100^{\circ}$ K, $200^{\circ}$ K, and $300^{\circ}$ K, 
shown by the red solid, green long-dashed and blue short-dashed lines, respectively, for (a) 
$a=100$ nm and (b) $a=1000$ nm.}~\label{c3deltat}
\end{figure}

Further, we show the variations in the $C_3$ coefficients by plotting them as function of the gap parameter $\Delta$ in 
Figs.~\ref{c3deltaa} and ~\ref{c3deltat}. In Fig.~\ref{c3deltaa}, we plot the dispersion coefficients for the Rb atom 
at four different values of the separation distances with varying $\Delta$ values from $10^{-4}$ eV (below which 
the $C_3$ coefficients are found to be insensitive) to 0.1 eV. From this graph, we find that the interactions depend on the gap 
parameters and the observed changes are almost in the factors of 10, 20, 7 and 4 (in percentage) of the $C_3$ values  
for the $a$ values of 10, 100, 500, and 1000 nm, respectively. We conclude from these observations that at 
the intermediate separation distances, the changes in the $C_3$ coefficients are maximum for the varying
values of the gap parameter. 

The calculated results for the $C_3$ coefficients, as functions of $\Delta$, for different temperatures are presented 
in Fig.~\ref{c3deltat}. In this figure, the lower solid line corresponds to the temperatures at $T=100^{\circ}$ K, 
the dashed line at $T=200^{\circ}$ K and the dotted line at $T=300^{\circ}$ K for two different `$a$' values. From 
Fig.~\ref{c3deltat}(a) with $a=100$ nm, we observe that the $C_3$ coefficients vary strongly with the gap parameter. 
Therefore, the region of intermediate distances are the ideal regime where comparison between the measured and 
calculated interaction potentials can offer to extract a suitable value for the gap parameter to describe the 
interactions of atoms with graphene more appropriately. Similarly from Fig.~~\ref{c3deltat}(b) with $a=1000$ nm, we find that 
(i) at $T=200^{\circ}$ K and $300^{\circ}$ K, the $C_3$ coefficients vary negligibly with the gap parameter, (ii) at 
$T=100^{\circ}$ K, the $C_3$ coefficients do not vary much up to $\Delta < 0.04$ eV with the gap parameter and (iii) 
at $T=100^{\circ}$ K, the $C_3$ coefficients vary appreciably for $\Delta > 0.04$ eV with the gap parameter. 
Therefore, we arrive at the conclusion from this study that at the larger distances, the region of intermediate 
temperatures are better suited to offer for the extraction of a more realistic value of the gap parameter. 

\subsection{$C_3$ as a function of temperature}
\begin{figure}[t]
\includegraphics[scale=0.67]{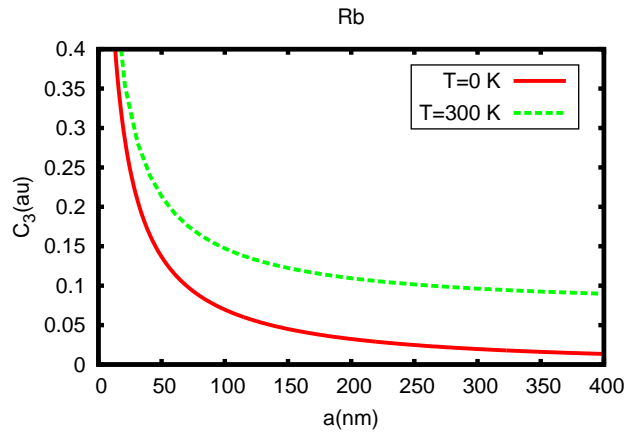}
\caption{(Color online) The $C_3$ coefficients (in au) as function of atom-graphene separation distance for the 
Rb atom at $T=0^{\circ}$ K and $T=300^{\circ}$ K. }~\label{temp}
\end{figure}

\begin{figure}[t]
\includegraphics[scale=0.67]{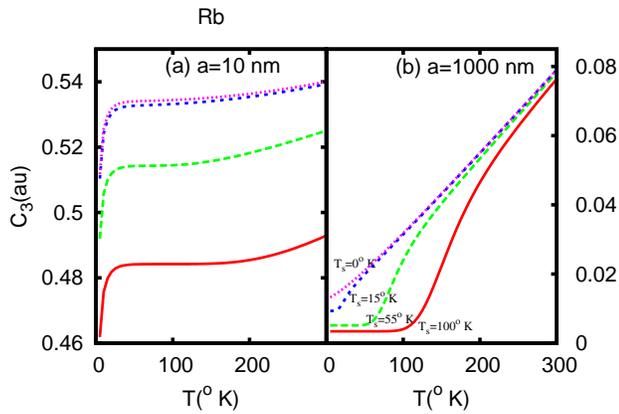}
\caption{(Color online) The $C_3$ coefficients (in au) as a function of temperature calculated at (a) $a=10$ nm, and (b) $a=1000$ nm, for different values of gap parameter where the 
lowest solid line corresponds to $\Delta=0.1$ eV, long-dashed line to $\Delta=0.05$ eV, short-dashed line to $\Delta=0.01$ eV and the uppermost dotted line represents
$\Delta=0.001$ eV. T$_s$ is the temperature upto which $C_3$ remains constant for a given $\Delta$ value.}~\label{c3td}
\end{figure}

To show the temperature dependencies on the $C_3$ coefficients, we only consider the interactions between the Rb atom 
and graphene which are more sensitive than the other atoms. In Fig.~\ref{temp}, we plot $C_3$ coefficients for the Rb 
atom as a function of the separation distance for two different temperatures; i.e. at $T=0^{\circ}$ K (solid line) and
$T=300^{\circ}$ K (dashed line). Results at the temperature $T=0^{\circ}$ K are obtained by using the Lifshitz theory 
for graphene-atom interaction as has been reported in our previous study \cite{harjeet}. From the figure, we observe 
appreciable differences in the results for different values of the temperature. These differences increase with the 
increasing values of the separation distance between the atom and the graphene layer. 

Next, we calculate the $C_3$ coefficients for the Rb atom as a function of temperature for four different values of 
the gap parameter at the separation distances $a=10$ nm (Fig.~\ref{c3td}(a)) and $a=1000$ nm (Fig.~\ref{c3td}(b)). In 
these figures, the solid line corresponds to $\Delta=0.1$ eV, the long dashed line to $\Delta=0.05$ eV, the short 
dashed line to $\Delta=0.01$ eV and the dotted line to $\Delta=0.001$ eV. From Fig.~\ref{c3td}(a), we notice that 
(i) for $T<30^{\circ}$ K, the $C_3$ coefficients vary by large amount with the change in the temperature, (ii) for 
$T>30^{\circ}$ K, the $C_3$ coefficients vary only negligibly with the change in the temperature and (iii) for a given 
temperature, the $C_3$ coefficients depends strongly on the chosen gap parameter value. However, the plots for 
the $C_3$ coefficients with $\Delta=0.01$ eV and $\Delta = 0.001$ eV almost overlap. Similarly from Fig.~\ref{c3td}(b), 
we observe that (i) the $C_3$ coefficients remain constant up to a certain critical temperature value, say $T_s$, 
for a given $\Delta$ parameter, (ii) the value of $T_s$ decreases with decreasing values of the $\Delta$ parameter, 
i.e., as shown in the figure, we obtain $T_s \approx 100^{\circ}$ K, $50^{\circ}$ K, $15^{\circ}$ K and $0^{\circ}$ K for the 
$\Delta$ values of 0.1 eV, 0.05 eV, 0.01 eV and 0.0001 eV, respectively, (iii) the $C_3$ coefficients have strong 
dependencies on the temperature after the critical value $T_s$ and (iv) for temperatures in the intermediate range 
(say $50<T<100^{\circ}$ K), the $C_3$ coefficients depend strongly on the $\Delta$ value.

Thus if measurements of the Rb atom and graphene interaction potentials can be carried out either at the small
separation distances at any given temperature or at the large separation distances and for the intermediate
values of the temperature, then these experimental data in comparison with the present theoretical results can 
be of utmost usefulness to find out a justifiable value for the gap parameter to describe the interactions of 
the atoms with graphene more applicably.

\section{Conclusion}

Summarizing our work, we have investigated the dispersion $C_3$ coefficients of the atom-graphene interactions for   
the alkali Li, Na, K and Rb atoms as functions of the separation distance, the gap parameter and the temperature 
by using accurate values of the dynamic polarizabilities of the atoms that were determined by us earlier and 
calculating the reflection coefficients in the Dirac model. We also made an attempt to identify the regime, where 
we recommend to conduct experiments to extract out realistic values of the gap parameter for describing the 
atom-graphene interactions more appropriately. This is an extension to our previous work on the determination of the 
$C_3$ coefficients for the interactions of the above atoms with graphene at the zero temperature to the non-zero 
thermal conditions. Our results computed at the room temperature can facilitate the experimentalists to apprehend 
the interactions between the considered alkali atoms with graphene better and can guide them to investigate the
relevant properties in the right direction.

\section*{Acknowledgement}
The work of B.A. is supported by CSIR grant no. 03(1268)/13/EMR-II, India. K.K. acknowledges the financial support 
from DST. B.K.S. acknowledges use of the PRL 3TFlop HPC cluster at Ahmedabad.

\bibliography{ref1.bib}

\end{document}